\input harvmac.tex
\vskip 2in
\Title{\vbox{\baselineskip12pt
\hbox to \hsize{\hfill}
\hbox to \hsize{\hfill}}}
{\vbox{\centerline{New 
Superstring Isometries and Hidden Dimensions}}}
\vskip 0.2in
{\vbox{\centerline{}}}
\centerline{Dimitri Polyakov\footnote{$^\dagger$}
{dp02@aub.edu.lb}}
\medskip
\centerline{\it Center for Advanced Mathematical Sciences}
\centerline{\it and  Department of Physics }
\centerline{\it  American University of Beirut}
\centerline{\it Beirut, Lebanon}
\vskip .4in
\centerline {\bf Abstract}
We study the hierarchy of 
hidden space-time symmetries
of noncritical strings in RNS formalism, realized nonlinearly.
Under these symmetry transformations the variation of the matter part of the
RNS action is cancelled by that of the ghost part.
These symmetries, referred to as the $\alpha$-symmetries,
 are induced by special space-time generators, violating
the equivalence of ghost pictures.
We classify the $\alpha$-symmetry generators 
in terms of superconformal ghost cohomologies $H_{n}\sim{H_{-n-2}}(n\geq{0})$
and associate these generators with a chain
of hidden space-time dimensions, with each ghost cohomology
$H_{n}\sim{H_{-n-2}}$ ``contributing'' an extra dimension.
Namely, we show that  
each ghost cohomology $H_{n}\sim{H_{-n-2}}$ 
of non-critical superstring theory in $d$-dimensions
contains
 $d+n+1$ $\alpha$-symmetry generators and 
the generators from   $H_{k}\sim{H_{-k-2}},1\leq{k}\leq{n}$,
combined together, extend the space-time isometry group
from the naive $SO(d,2)$ to $SO(d+n,2)$.
In the simplest case of $n=1$ the $\alpha$-generators are identified
with the extra symmetries of the $2T$-physics formalism, also known
to originate from a hidden space-time dimension.
 {\bf }
\Date{June 2007}

\vfill\eject
\lref\verl{H. Verlinde, Phys.Lett. B192:95(1987)}
\lref\bars{I. Bars, Phys. Rev. D59:045019(1999)}
\lref\barss{I. Bars, C. Deliduman, D. Minic, Phys.Rev.D59:125004(1999)}
\lref\barsss{I. Bars, C. Deliduman, D. Minic, Phys.Lett.B457:275-284(1999)}
\lref\barsf{C. Vafa, Nucl. Phys. B469:403-418 (1996)}
\lref\barsff{I. Bars, Phys. Rev. D55:2373-2381 (1997)}
\lref\lian{B. Lian, G. Zuckerman, Phys.Lett. B254 (1991) 417}
\lref\pol{I. Klebanov, A. M. Polyakov, Mod.Phys.Lett.A6:3273-3281}
\lref\wit{E. Witten, Nucl.Phys.B373:187-213  (1992)}
\lref\grig{M. Grigorescu, math-ph/0007033, Stud. Cercetari Fiz.36:3 (1984)}
\lref\witten{E. Witten,hep-th/0312171, Commun. Math. Phys.252:189  (2004)}
\lref\wb{N. Berkovits, E. Witten, hep-th/0406051, JHEP 0408:009 (2004)}
\lref\zam{A. Zamolodchikov and Al. Zamolodchikov,
Nucl.Phys.B477, 577 (1996)}
\lref\mars{J. Marsden, A. Weinstein, Physica 7D (1983) 305-323}
\lref\arnold{V. I. Arnold,''Geometrie Differentielle des Groupes de Lie'',
Ann. Inst. Fourier, Grenoble 16, 1 (1966),319-361}
\lref\self{D. Polyakov,  Int. Jour. Mod. Phys A20: 2603-2624 (2005)}
\lref\selff{D. Polyakov, Phys. Rev. D65: 084041 (2002)} 
\lref\sellf{D. Polyakov, Int. J. Mod. Phys A20:4001-4020 (2005)}
\lref\selfian{I.I. Kogan, D. Polyakov, Int.J.Mod.Phys.A18:1827(2003)}
\lref\doug{M.Douglas et.al. , hep-th/0307195}
\lref\dorn{H. Dorn, H. J. Otto, Nucl. Phys. B429,375 (1994)}
\lref\prakash{J. S. Prakash, 
H. S. Sharatchandra, J. Math. Phys.37:6530-6569 (1996)}
\lref\dress{I. R. Klebanov, I. I. Kogan, A. M.Polyakov,
Phys. Rev. Lett.71:3243-3246 (1993)}
\lref\selfdisc{ D. Polyakov, hep-th/0602209, 
Int. J. Mod.Phys.A22:1375-1394 (2007)}
\lref\selfgc{D. Polyakov, hep-th/0701044,
to appear in IJMPA}

\centerline{\bf 1. Introduction}

In our recent work ~{\selfgc} we have shown that non-critical
RNS  superstring theories are invariant under the  set of unusual 
nonlinear space-time transformations, not at all
evident from the structure of their worldsheet actions.
That is, consider the worldsheet 
action of $d$-dimensional non-critical RNS superstring theory
 given by
\eqn\grav{\eqalign{S={1\over{2\pi}}\int{d^2z}\lbrace
-{1\over2}{\partial{X_m}}\bar\partial{X^m}-{1\over2}
\psi_m\bar\partial\psi^m-{1\over2}\bar\psi_m\partial\bar\psi^m\rbrace
+S_{ghost}+S_{Liouville}\cr
S_{ghost}={1\over{2\pi}}\int{d^2z}\lbrace{b}\bar\partial{c}+{\bar{b}}
\partial{\bar{c}}+
\beta\bar\partial\gamma+\bar\beta\partial\bar\gamma\rbrace
\cr
S_{Liouville}={1\over{4\pi}}\int{d^2z}
{\lbrace}\partial\varphi\bar\partial\varphi+\lambda\bar\partial\lambda+
{\bar\lambda}\partial\bar\lambda-F^2+2\mu_0be^{b\varphi}
(ib\lambda\bar\lambda-F)\rbrace
}}
where $\varphi,\lambda,F$ are the components of the
 Liouville superfield, $X^m (m=0,...,d-1)$
are the space-time coordinates,$\psi^m,\bar\psi^m$ are
their worldsheet superpartners; $b,c,\beta,\gamma$
are the fermionic and bosonic (super)reparametrization ghosts bosonized as
\eqn\grav{\eqalign{b=e^{-\sigma};c=e^\sigma,\cr
\beta=e^{\chi-\phi}\partial\chi\equiv\partial\xi{e^{-\phi}};
\gamma=e^{\phi-\chi}.\cr
}}.

This action is obviously invariant under two global
$d$-dimensional space time symmetries -
Lorenz rotations and translations. 
 One can straightforwardly  check, however, that in addition to these 
obvious symmetries
the action (1) is also invariant under the following nonlinear
global transformations, mixing the matter and the ghost sectors of the
theory ~{\selfgc}:

\eqn\grav{\eqalign{\delta{X^m}=\epsilon
\lbrace\partial(e^\phi\psi^m)+2e^\phi\partial\psi^m\rbrace\cr
\delta\psi^m=\epsilon\lbrace{-}e^\phi\partial^2{X^m}
-2\partial(e^\phi\partial{X^m})\rbrace\cr
\delta\gamma=\epsilon{e^{2\phi-\chi}}(\psi_m\partial^2{X^m}
-2\partial\psi_m\partial{X^m})\cr
\delta\beta=\delta{b}=\delta{c}=0}}

The
 variations of the matter and the ghost parts
of the RNS action (1) under the transformations (3) are given by

\eqn\grav{\eqalign{\delta_{S_{matter}}=-{{\epsilon}\over{2\pi}}
\int{d^2z}(\bar\partial{e^\phi})\lbrace\partial^2{X_m}\psi^m-2\partial{X_m}
\partial\psi^m\rbrace\cr
\delta{S_{ghost}}\equiv\delta{S_{\beta\gamma}}=-\delta{S_{matter}}}}

so the action is symmetric under (3), as the variation of the matter part
is cancelled by that of the ghost part.

It is easy to check that the generator of the transformations
 (3)  is given by
\eqn\lowen{T=\int{{dz}\over{2i\pi}}{e^\phi}(\partial^2{X_m}\psi^m
-2\partial{X_m}\partial\psi^m)}

The integrand of (5) is a primary field of dimension 1,
i.e. a physical generator.
While it is not manifestly BRST-invariant (it
doesn't commute with the supercurrent terms of $Q_{brst}$),
 its BRST invariance can be restored by 
adding the appropriate b-c ghost
dependent  terms, according to the prescription described in 
~{\selfgc}.
The peculiar property of this generator is that 
it is annihilated by $\Gamma^{-1}=c\partial\xi{e^{-2\phi}}$ and
 has no analogues
at higher pictures, such as $0,-1$ and $-2$
(but has versions at higher positive pictures $+2,+3,...$
which can be obtained by  
using the direct picture-changing
operator $\Gamma=\lbrace{Q_{brst}},\xi\rbrace$). 
The physical operators with such a property
are referred to as the  positive ghost number $+1$ cohomology $H_1$,
or simply ghost cohomology 1 ~{\selfgc}.

There is however, a picture $-3$ version of this
generator, with the manifest BRST invariance. This version can be 
obtained simply by replacing $e^\phi\rightarrow{e^{-3\phi}}$ in (5).
Similarly to the picture $+1$-version, the picture ${-3}$
version is annihilated by $\Gamma$, so there are no versions of this
operator at pictures $-2,-1$ and 0 while the versions at pictures
below $-3$ can be obtained by straightforward inverse picture changing.
For this reason, the picture $-3$ version of (5) is an element
of negative ghost cohomology number $H_{-3}$.
 For the sake of completeness, let us recall some definitions
made in ~{\selfgc}.
 The positive ghost cohomologies
$H_{n}(n\geq{1})$
consist of physical operators existing at positive ghost pictures
greater or equal to $n$, annihilated
by $\Gamma^{-1}$ at the minimal positive picture $n$.
That is, the picture changing transformations (direct and inverse)
allow to move the elements of $H_n$ between  pictures
greater or equal to $n$, but not below $n$.
Similarly, the negative ghost cohomologies, $H_{-n} (n\geq{3})$
consist of physical operators existing at 
 ghost pictures $-n$ and below,
while at their minimal negative picture $-n$ they are annihilated
by the direct picture-changing $\Gamma$.

The zero cohomology $H_0$ by definition consists of picture-independent
operators (i.e. standard perturbative vertex operators
 existing at all the ghost pictures). The cohomologies
$H_{-1}$ and $H_{-2}$ are empty, since any operator
at picture $-1$ or $-2$ is either BRST-exact or can be 
transformed to picture zero by $\Gamma$.
The question of  existence of nonzero ghost cohomologies
(which by itself is quite non-trivial) has been discussed in details
previously ~{\selfgc}. The important property of the ghost
cohomologies is the isomorphism between positive and negative H's:
$$H_{n}\sim{H_{-n-2}}$$ constructed in ~{\selfgc}.
This isomorphism is $not$ a picture changing equivalence, though it
is somewhat reminiscent of it. 
Structurally, it maps cohomologies of positive and
 negative numbers, ``bypassing'' the  pictures in the middle.
Typically,  commutators of the current algebra of
 operators from nonzero H's involve the currents from both
negative and positive  cohomologies, related by the isomorphism.

It is not difficult to show
 that, just like the operator (5) - the element of $H_{1}$ -
generates the space-time symmetry transformations (3) of the RNS action (1),
its picture $-3$ version (the element of $H_{-3}$) generates the 
symmetry transformations identical to (3),
with $e^\phi$ replaced by $e^{-3\phi}$.
For the critical ($d=10$) uncompactified
RNS superstrings the set of the transformations (3) is the only
additional nonlinear space-time symmetry.
For non-critical strings ($d\neq{10}$), however, there are
$d+1$ additional $\alpha$-symmetries, involving the Liouville sector.
The appropriate transformations are given by 

\eqn\grav{\eqalign{\delta{X_m}=\epsilon_{m\alpha}{\lbrace}
\partial(e^\phi\lambda)+2e^{\phi}\partial\lambda
\rbrace\cr
\delta\lambda
=-\epsilon_{m\alpha}{\lbrace}
2\partial(e^\phi\partial{X^m})+e^\phi
\partial^2{X^m}\rbrace\cr
\delta\gamma=\epsilon_{m\alpha}e^{2\phi-\chi}\lbrace
\partial^2{X^m}\lambda
-2\partial{X^m}\partial\lambda\rbrace\cr
\delta\beta=\delta{b}=\delta{c}=\delta\varphi=\delta\psi^m=0}}

and

\eqn\grav{\eqalign{\delta\varphi
=\epsilon_{-\alpha}\lbrace
\partial(e^\phi\lambda)+2e^\phi\partial\lambda
\rbrace\cr
\delta\lambda=-\epsilon_{-\alpha}\lbrace{2}\partial(e^\phi\partial\varphi)
+e^\phi{\partial^2}\varphi\rbrace\cr
\delta\gamma=\epsilon_{-\alpha}e^{2\phi-\chi}\lbrace
\lambda\partial^2\varphi-2\partial\varphi\partial\lambda
\rbrace\cr
\delta\beta=\delta{b}=\delta{c}=\delta{X^m}=\delta\psi^m=0}}.

In the limit of zero cosmological constant,
the dressed generators inducing the transformations (6) and (7) are given by
~{\selfgc}
\eqn\grav{\eqalign{L^{m\alpha}=l(d)\oint{{dz}\over{2i\pi}}
e^{\phi+Q\varphi}\lbrace(\partial^2\varphi+Q(\partial\varphi)^2)\psi^m
-2\partial\varphi\partial\psi^m+\partial^2{X^m}\lambda\cr
-2\partial{X^m}
\partial\lambda-4Q\partial\varphi\lambda\partial{X^m}\rbrace}}
and
\eqn\lowen{L^{-\alpha}=l(d)\oint{{dz}\over{2i\pi}}e^{\phi+Q\varphi}
\lbrace(\partial^2\varphi+Q(\partial\varphi)^2)\lambda+
\partial\varphi\partial\lambda\rbrace}
accordingly with the normalization constant
$l(d)$ given by ~{\dress,\selfgc}

\eqn\lowen{l(d)={\sqrt{{k(d)+1}\over{k(d)+2}}}}
with
\eqn\lowen{k(d)={1\over{8}}(d-1\pm{\sqrt{(d-1)(d-9)}})}

The generators (8) and (9) are the Virasoro primaries, annihilated
by the inverse picture changing. They are BRST non-trivial and invariant
(upon adding the $b-c$ ghost correction terms which we have skipped)
 and therefore
are the elements of $H_1\sim{H_{-3}}$. As before, the $H_{-3}$ version
of the generators (8), (9) with the manifest
BRST-invariance (with no $b-c$ correction terms)
and the set of the related space-time transformations
can be obtained simply by replacing
$\phi\rightarrow{-3\phi}$ in (6) - (9).
Combined with ${{(d+1)(d+2)}\over2}$ dimension 1 Virasoro primaries:
\eqn\grav{\eqalign{L^{mn}=\oint{{dz}\over{2i\pi}}\psi^{m}{\psi^n}\cr
L^{+m}=\oint{{dz}\over{2i\pi}}e^{-\phi}\psi^m\cr
L^{-m}=l(d)\oint{{dz}\over{2i\pi}}e^{Q\varphi}\psi^m\lambda\cr
L^{++}=l(d)\oint{{dz}\over{2i\pi}}e^{-\phi+Q\varphi}\lambda,}}

that induce $d+1$ translations and ${{d(d+1)}\over2}$
rotations in space-time (including the Liouville direction),
the $d+2$ currents (5), (8), (9) of $H_1\sim{H_{-3}}$ 
 enlarge the current algebra of the space-time symmetry generators from
$SO(d,2)$ to $SO(d+1,2)$, effectively bringing in extra 
dimension to the theory.
Indeed,
introducing the $d+2$-dimensional index
$M=(m,+,-,\alpha);m=0,...{d-2};\alpha=1$ with the $(d,2)$ metric
$\eta^{MN}$ consisting of $\eta^{mn},\eta^{+-}=-1,
\eta^{--}=\eta{++}=0,\eta^{\alpha\alpha}=1$
and evaluating the commutators of the operators (5), (8),
(9), (12) it isn't difficult
to check that ~{\selfgc}
\eqn\lowen{\lbrack{L^{M_1N_1},L^{M_2N_2}\rbrack}
=\eta^{M_1M_2}L^{N_1N_2}+\eta^{N_1N_2}L^{M_1M_2}
-\eta^{M_1N_2}L^{M_2N_1}-\eta^{M_2N_1}L^{M_1N_2}}

Note that, as the $SO(d,2)$ 
space-time symmetry 
group of translations and rotations for
 non-critical RNS strings is identical to the isometry
group of the $AdS_d$ space,  the $H_{1}\sim{H_{-3}}$ generators (5), (8),
(9)
of the lowest nonzero ghost cohomology are simply the stringy
analogues of the off-shell symmetry generators from hidden
space-time  dimension, observed in the 2T physics approach
~{\bars, \barss, \barsss} for  a particle in the $AdS_d$ space 
~{\bars}.
The precise correspondence between the
space-time symmetry 
generators for a $AdS_d$ particle in the 2T physics formalism 
(including the extra generators from hidden dimension)
and the space-time symmetry generators for 
non-critical strings(including
$H_1\sim{H_{-3}}$
$L$-generators of $\alpha$-symmetries ) is
given by
\eqn\grav{\eqalign{
L^{mn}\leftrightarrow{T^{mn}}\cr
L^{+m}\leftrightarrow{T^{+m}},
L^{-m}\leftrightarrow{T^{-m}}\cr
L^{++}\leftrightarrow{T^{+-}}}}
for the ``standard'' generators of the $SO(d,2)$ subgroup
of $SO(d,2)$
and
\eqn\grav{\eqalign{L^{+\alpha}\leftrightarrow{T^{+}},
L^{-\alpha}\leftrightarrow{T^{-}}\cr
L^{m\alpha}\leftrightarrow{T^{m}}}}
for the extra generators (associated with the higher space-time dimension).
 Here the symmetry generators for the $AdS_d$ particle are given by
(using the notations of ~{\bars}):
\eqn\grav{\eqalign{
T^{mn}=x^mp^n-x^np^m,T^{+m}=p^m,T^{+-}=-uk+({\vec{x}}{\vec{p}})\cr
T^{-m}={{p^m}\over{2u^2}}+uk{x^m}+{1\over2}x^2p^m-({\vec{x}}{\vec{p}})x^m\cr
T^{-}={1\over2}k+{1\over2}x^2(u^2k-2ku^2)-{{({\vec{x}}{\vec{p}})}\over{u}}\cr
T^{+}=-u^2k-2ku^2,T^m=x^m(u^2k-2ku^2)-{{p^m}\over{u}}}}
where the $AdS_d$  metric is given by
\eqn\grav{\eqalign{ds^2=u^2(dx^m)^2+{{du}\over{u^2}}\cr
m=0,1,...,d-2
}}
and $p^m$ and $k$ are the canonical conjugates
of $x^m$ and $u$ respectively.
Here $u$ is the radial $AdS_d$ coordinate, corresponding to the Liouville
direction on the string theory side.
Thus the picture-dependent generators of $H_1\sim{H_{-3}}$
of the lowest nonzero ghost cohomology are in one to one correspondence
with the symmetry generators from hidden space-time dimension in the $2T$
approach. In the following section we shall discuss the generalization
of this result to ghost cohomologies of higher orders.

\centerline{\bf 2. $\alpha$-Symmetries at the level $H_{2}\sim{H_{-4}}$}

The natural question is how to extend the results, described
in the previous section,
to higher order cohomologies, i.e. the BRST-invariant primaries from $H_2\sim
{H_{-4}},H_3\sim{H_{-5}}$, etc.
Do these currents generate more space-time symmetries?
If yes, are these new symmetries related to yet unknown higher space-time
dimensions, not detected in the $2T$ formalism?
Can these extra dimensions be understood in terms of
the class of holography relations, so that each extra dimension
is effectively  generated by operators from the associate ghost cohomology?

In the rest of the paper, we will try to address these questions.
It is instructive to start from the
  simplest example of non-critical RNS strings - the supersymmetric 
$c=1$ model where the straightforward
construction of the elements of
 $H_n\sim{H_{-n-2}}$  is relatively simple.
It has been shown ~{\selfdisc} that the 
physical operators from the higher ghost cohomologies
enlarge the current algebra (which becomes the target space symmetry
of the theory when the space dimension is compactified at the
self-dual radius).That is, while the original current algebra
of the theory is given by SU(2) ~{\lian, \pol, \wit},
so the standard picture-independent
discrete states are the SU(2) multiplets, introducing the
target space symmetry generators  of higher cohomologies of ghost numbers
up to $N$ (i.e. extending the current algebra  with the elements of
 $H_n\sim{H_{-n-2}}$ with $n=1,2,...N$) enhances the symmetry
algebra of the $c=1$ theory (compactified at self-dual radius)
from $SU(2)$ to $SU(N+2)$.
 Let us review  first the simplest case of $N=1$.
One starts with the $H_1\sim{H_{-3}}$ generator 
\eqn\lowen{T_{-3,2}=\oint{{dz}\over{2i\pi}}e^{-3\phi+2iX}\psi}
with momentum $+2$ in the $X$-direction
(the left index refers to the ghost cohomology number and the right
one to the momentum)
and acts on it repeatedly with the lowering operator
$T_{0,-1}=\oint{{dz}\over{2i\pi}}e^{-iX}\psi$ of $SU(2)$,
obtaining altogether five $H_{-3}\sim{H_1}$ generators with the 
discrete momenta$-2\leq{p}\leq{2}$ (note that the momentum $-2$ generator 
is annihilated by
$T_{0,-1}$).
Unified with  3 currents of $SU(2)$, 5 currents of $H_1\sim{H_{-3}}$
combine into $8$ operators generating $SU(3)$ ~{\selfdisc}.
The lowering subalgebra of $SU(3)$ consists of 3 operators with 
negative discrete momenta (one from $H_0$ and two from $H_1\sim{H_{-3}}$),
the upper subalgebra includes those with the positive momenta
while two zero momentum generators 
$T_{00}\sim{\oint{{dz}\over{2i\pi}}}\partial{X}$ of $H_0$ and
\eqn\lowen{T_{-3,0}=\oint{{dz}\over{2i\pi}}e^{-3\phi}(\partial^2{X}\psi-
2\partial{X}\partial\psi)}
generate the Cartan subalgebra.
This construction also can be  generalized for 
the case of arbitrary $H_N\sim{H_{-N-2}}$ ~{\selfdisc}.
To construct the generators of $H_N\sim{H_{-N-2}}$ one starts with
$T_{-N-2,N+1}=\oint{{dz}\over{2i\pi}}e^{-(N+2)\phi+i(N+1)X}\psi$
and acts repeatedly with $T_{0,-1}$, obtaining $2N+3$ generators
with the discrete momenta $-N-1\leq{p}\leq{N+1}$.
Unifying all the operators of $H_n\sim{H_{-n-2}},0\leq{n}\leq{N}$,
obtained by this procedure, one
combines them into $(N+2)^2-1$ generators of  $SU(N+2)$.
Among these generators there are $N+1$ currents with the zero momentum
that
form the Cartan subalgebra of $SU(N+2)$ (all the Cartan generators 
appear from different ghost cohomologies).
 Acting with the
lowering generators of $SU(N+2)$ on the highest weight 
vectors (which typically are the tachyonic primaries at integer momenta)
one  generates the extended physical spectrum of picture-dependent
physical states - the SU(N+2) multiplets. This has been shown explicitly
for $N=1$ and $N=2$ and conjectured for higher values of $N$. Accordingly,
the structure constants of the vertex operators are given by
the $SU(N+2)$ Clebsch-Gordan coefficients and thus the operator
algebra is isomorphic to the volume-preserving diffeomorphisms in $N+2$
dimensions ~{\selfdisc, \pol}.
 The natural interpretation of this result
is that each separate nonzero ghost cohomology brings in an effective
extra dimension and this
implies the holographic correspondence between $c=1$ supersymmetric model,
including the states from nonzero $H_{n}$,  to 
field theories in higher dimensions.
As we have seen before, higher dimensional RNS models
follow the same pattern for $N=1$, as the currents 
 (5), (8), (9) of $H_1\sim{H_{-3}}$
increase the 
dimensionality of the space-time symmetry group by 1 unit.
There is, however, an important difference. The case of $c=1$ compactified
at the self-dual radius, is special as the symmetry generators
include operators at nonzero momenta. The uncompactified
non-critical strings in $d>1$, however, 
cannot contain generators at non-zero momenta since
in operators of the type $\sim{e^{ikX}}$ the left and right modes aren't
separated and such generators generally cannot be integrated over a
 worldsheet contour.
For this reason, the only 
 $H_{n}\sim{H_{-n-2}}$
generators that can be exported
from the $c=1$ case to higher dimensions are those with zero momentum, i.e.
the Cartan generators of $SU(n+2)$.  For example, the 
Liouville-independent part of the $H_1\sim{H_{-3}}$ 
generator of  $\alpha$-symmetry (5) 
has the structure identical to the Cartan of $SU(3)$
(that can be interpreted as a ``hypercharge operator'' ~{\selfdisc})
For large values of $n$ the expressions for Cartan
generators of $SU(n+2)$ are more complicated ~{\selfdisc}, 
however, the relation to Cartan generators is still useful
in order to to conjecture some properties of
$H_{n}\sim{H_{-n-2}}$ generators to simplify our searches.
In particular, it is quite clear that 

1)expressions for the $H_{n}\sim{H_{-n-2}}$
generators do not contain any derivatives of the superconformal ghost
field $\phi$, so their ghost dependence is determined
entirely by $e^{n\phi}$ or $e^{-(n+2)\phi}$

2) the matter parts of $SU(n)$ Cartan generators of $c=1$
can be used as building blocks to construct the higher cohomolody elements
for non-critical strings in higher space-time dimensions.

Given these indications, below we shall look for the elements
of $H_{2}\sim{H_{-4}}$ generating the higher order $\alpha$-transformations.
To avoid technical complications with inverse picture-changing
due to the $b-c$ ghost terms in the $H_{2}$ representation, we shall
concentrate on
the $H_{-4}$-version of the generators.
We start from the Liouville-independent part of the 
$\alpha$-transformations (i.e. those that mixes matter with ghosts
but doesn't touch the Liouville sector).
In general, these cohomology elements - the candidates for the 
$\alpha$-symmetry generators - could be in various tensor representations
of the Lorenz group, so we start with the simplest case of
the scalar generator.
The most general expression for a scalar dimension 1 operator of 
the ghost number
$-4$, in view of the above conditions, is given by
\eqn\grav{\eqalign{
V(z)=e^{-4\phi}\lbrace{\alpha_1}(\partial{X_m}\partial{X^m})(\partial{X^n}
\partial^2{X^n})+\alpha_2(\partial{X_m}\partial{X^m})(\psi_n\partial^2\psi^n)
\cr
+\alpha_3(\partial{X_m}\partial^2{X^m})(\psi_n\partial\psi^n)
+\alpha_4(\psi_m\partial\psi^m)(\psi_n\partial^2\psi^n)
+\beta_1(\psi_m\partial{X^m})(\partial{X_n}\partial^2\psi^n)
\cr
+\beta_2(\psi_m\partial{X^m})(\partial^2{X_n}\partial\psi^n)
+\beta_3(\psi_m\partial{X^m})(\psi_n\partial^3{X^n})
+\beta_4(\psi_m\partial^2{X^m})(\partial\psi_n\partial{X^n})
\cr
+\gamma_1(\partial{X_m}\partial^4{X^m})+\gamma_2(\partial^2{X_m}
\partial^3{X^m})+\lambda_1(\psi_m\partial^4\psi^m)+\lambda_2
(\partial\psi_m\partial^3\psi^m)\rbrace(z)}}
where $\alpha_i,\beta_i,\gamma_i$ and $\lambda_i$ are some numbers.
In order to be  the element of $H_{-4}$, $V(z)$ of (20) must satisfy
two necessary conditions:

1) it must be a primary field, i.e. its $OPE$ with the full matter$+$ghost
stress-energy tensor must not contain singularities of the order higher than
$(z-w)^{-2}$, i.e. all the OPE coefficients in front of the terms 
of the order of $(z-w)^{-n},n\geq{3}$ have to vanish separately.

2) it must be annihilated by the picture-changing operator,
i.e. its OPE with $\Gamma$ can only contain terms of the order of
$(z-w)^{n},n\geq{1}$, that is, the coefficients in front of the lower order
terms must all vanish separately.

These two conditions altogether generate a number of linear constraints
on the  $\alpha_i,\beta_i,\gamma_i$ and$\lambda_i$ coefficients.
Any combination of the coefficients, satisfying these constraints
(in addition to the BRST triviality condition), is
a candidate  for the element of $H_{-4}$ and
for the higher order $\alpha$-symmetry generator.
Note that the primary field $+$ annihilation constraints
are necessary, but generally not sufficient conditions
to define the $\alpha$-generator since the latter
also has to be BRST-nontrivial, as BRST-exact currents obviously
can't generate any new global space-time symmetries.
 The non-triviality needs be checked separately or
alternatively, one has to verify straightforwardly
that the transformations
induced by the candidate operator, are indeed the symmetries
of the RNS action.

We start with the primary field constraints on $V$.
Using the matter$+$ghost stress-energy tensor:
\eqn\grav{\eqalign{
T(z)=-{1\over2}\partial{X_m}\partial{X^m}-{1\over2}\partial\psi_m\psi^m
-{1\over2}(\partial\varphi)^2+{Q\over2}\partial^2\varphi
+T_{b-c}+T_{\beta\gamma}}}

the straightforward calculation gives

\eqn\grav{\eqalign{T(z)V(w)=
(z-w)^{-7}(48\lambda_1+6\lambda_2-24\gamma_1-12\gamma_2)
\cr
+(z-w)^{-5}\lbrace(2\beta_2+2d\alpha_3-12\lambda_1+12\lambda_2
+2(1-d)\alpha_4)\partial\psi_m\psi^m\cr
+(2d\alpha_2-2d\alpha_1+2\beta_1+24\gamma_1)\partial{X_m}\partial{X^m}
\rbrace\cr
+(z-w)^{-4}\lbrace({1\over2}(1-d)\alpha_4+\beta_1+d\alpha_2)
\partial^2\psi_m\psi^m\cr
+({1\over2}(d\alpha_3+\alpha_4+\beta_2+\beta_4)
-(d+2)\alpha_1+6\gamma_2)\partial{X_m}\partial^2{X^m}\rbrace\cr
\cr
+(z-w)^{-3}\lbrace
\lambda_2\psi_m\partial^3\psi^m+2\gamma_2(\partial{X_m}\partial^3{X^m})
+2\alpha_1(\partial{X_m}\partial{X^m})^2
\cr
+\alpha_4
(\partial\psi_m\psi^m)^2+(2\alpha_3+\alpha_2)(\partial{X_m}\partial{X^m})
(\psi_m\partial\psi^m)\cr
+(\beta_1+2\beta_2+2\beta_4)(\psi_m\partial{X^m})
(\partial\psi_m\partial{X^m})
\cr
+(\beta_2-\beta_4)(\psi_m\partial{X^m})
(\psi_m\partial^2{X^m})\rbrace+O((z-w)^{-2})}},

so the primary field constraints are

\eqn\grav{\eqalign{
48\lambda_1+6\lambda_2-24\gamma_1-12\gamma_2=0\cr
2\beta_2+2d\alpha_3-12\lambda_1+12\lambda_2
+2(1-d)\alpha_4=0\cr
2d\alpha_2-2d\alpha_1+2\beta_1+24\gamma_1=0\cr
{1\over2}(1-d)\alpha_4+\beta_1+d\alpha_2=0\cr
{1\over2}(d\alpha_3+\alpha_4+\beta_2+\beta_4)
-(d+2)\alpha_1+6\gamma_2=0\cr
\lambda_2=\gamma_2=\alpha_1=\alpha_4=0\cr
2\alpha_3+\alpha_2=0\cr
\beta_1+2\beta_2+2\beta_4=0\cr
\beta_2-\beta_4=0}}

Next, consider the annihilation condition  $\Gamma{V}{\sim0}$.
Since $V$ is the dimension 1 integrand of vertex operator in the integrated
form, the ${c}\partial\xi$ term of $\Gamma$ of fermionic
ghost number $1$ doesn't act on $V$ while the  $b-c$ ghost number $-1$
term $\sim{b}e^{2\phi-\chi}(\partial\chi+\partial\sigma)$
annihilates $V$ for any choice of the coefficients $\alpha,\beta,\gamma$
and $\lambda$.
Therefore it is sufficient to consider how $V$ is acted on 
by the $b-c$ ghost number zero term $\sim{e^\phi}G_{matter}\sim{e^{\phi}}
\psi_m\partial{X^m}$. Note that, as the expression for $V$
is Liouville-independent, the OPE of $V$ with $e^{\phi}G_{Liouville}$
vanishes as well. The OPE evaluation of $\Gamma$ and $V$
gives

\eqn\grav{\eqalign{{e^\phi}\psi_m\partial{X^m}(z)V(w)
=((z-w)^{-1}+\partial\phi(z-w)^0)e^{-3\phi}\psi_m\partial{X^m}(w)
\cr\times{\lbrace}
-4\alpha_2-2\alpha_3-2(d+1)\beta_1-2d\beta_2+6(1-d)\beta_3-2\beta_4
-24\gamma_1+24\lambda_1\rbrace
\cr+(z-w)^0\lbrace
e^{-4\phi}\partial{X_m}\partial\psi^m(w)
(2\alpha_3+2\beta_2+2d\beta_4+6\lambda_2-24\gamma_1)\cr
+e^{-4\phi}\partial^2{X_m}\psi^m(w){\lbrace}-2\alpha_3-\beta_2-d\beta_4
-6\gamma_2+24\gamma_1{\rbrace}+O(z-w)}}

accordingly the annihilation constraints are
\eqn\lowen{\eqalign{
-4\alpha_2-2\alpha_3-2(d+1)\beta_1-2d\beta_2+6(1-d)\beta_3-2\beta_4
-24\gamma_1+24\lambda_1=0\cr
2\alpha_3+2\beta_2+2d{\beta}_4+6\lambda_2-24\gamma_1=0\cr
-2\alpha_3-\beta_2-d\beta_4
-6\gamma_2+24\gamma_1=0}}

Primary field constraints (22) along with the annihilation constraints (23)
define the  elements of the cohomology (provided
they are BRST non-trivial). Simple check
shows that the system of linear equations $(22),(23)$ has no
nonzero solutions, therefore $H_{-4}\sim{H_2}$ does not contain
any Liouville-independent scalars.

Now consider the Liouville-independent vector candidates for $H_{-4}$.
The analysis, presented in the Appendix of this paper,
 is analogous to the scalar case
and, as before,  we find no Liouville-independent
 BRST non-trivial vector operators
of $H_{-4}\sim{H_2}$
generating any new symmetries of the RNS action.
Similarly,  one can show that $H_{-4}\sim{H_2}$ contains
no higher rank tensors generating symmetries in the space-time.
Therefore, unlike the $H_{-3}\sim{H_1}$ case
(with one Liouville-independent generator), all
the  $\alpha$-symmetry generators of $H_2\sim{H_{-4}}$
 depend on the Liouville mode. Our aim is now to find
the Liouville-dependent currents of $H_2\sim{H_{-4}}$.
Instead of the straightforward procedure described above
(i.e. solving the annihilatilation $+$ primary field constraints
for a general picture $-4$ operator), involving lengthy calculations,
it is easier to guess the structure of the 
cohomology elements and then to verify
directly
that they satisfy the operator algebra generating a space-time symmetry
with an extra dimension and induce the space-time symmetry transformation.
For simplicity, in the rest of the paper
we will restrict ourselves to the case of the zero dilaton field
so we can neglect the background charge and the related effects  of 
 the second derivative term in the Liouville stress tensor.
In principle all our calculations can be straightforwardly
generalized to the case of non-zero dilaton as well
to account for the effect of the dressing (as
it has been done in (8), (9), (12) for the case of $H_1\sim{H_{-3}}$),
however the relevant expressions for the generators
become quite cumbersome and such complications aren't necessary
for our purposes.

With the dimension ${5\over2}$ scalar primary fields
\eqn\grav{\eqalign{F_1(X,\psi)=\partial^2{X_m}\psi^m-2\partial{X_m}
\partial\psi^m\cr
F_1(\varphi,\lambda)=\partial^2\varphi\lambda-2\partial\varphi\partial\lambda}}
it is not difficult to check that the generator
\eqn\grav{T=\oint{{dz}\over{2i\pi}}e^{-4\phi}
(\partial^2{X_m}\psi^m-2\partial{X_m}
\partial\psi^m)(\partial^2\varphi\lambda-2\partial\varphi\partial\lambda)}
is in the cohomology, i.e. is the scalar element of $H_{-4}$.
Note that $F_1(X,\psi)$ has the structure of the matter
part of the Cartan generator of $SU(3)$ (27) while $F_1(\varphi,\lambda)$
is its Liouville copy, with the $c=1$ matter fields $X$ and $\psi$
replaced with $\varphi$ and $\lambda$.
Similarly to (13), it is now appropriate to introduce
the $d+4$-dimensional space-time index
$M=(m,+,-.\alpha,\beta)$ with $m=0,...d-1;\alpha,\beta=1$, and
the $(m,+,-)$ indices correspond to $SO(d,2)$ isometry group
of $AdS_{d+1}$ generated by the matter$+$Liouville fields.
As before, the index $\alpha=1$ labels the extra space-time dimension generated
by the currents of $H_{-3}\sim{H_1}$ (that enlarge the
$AdS_{d+1}$ isometry with the $\alpha$-symmetry;
as the new index $\beta=1$, it
 will correspond to the
new extra dimension induced by $H_{-4}\sim{H_{2}}$ currents
of the higher order $\alpha$-symmetry.
As we expect the complete space-time symmetry group 
to be extended from $SO(d+1,2)$ to $SO(d+2,2)$ by the operators
of $H_2\sim{H_{-4}}$, we look for
$d+3$ generators $L^{m\beta},L^{\beta{+}},L^{\beta{-}}$
and $L^{\alpha\beta}$ of $H_{-4}\sim{H_2}$, i.e. for $1$ space-time
$d$-vector and 3 scalars.
The first step is to correctly identify the operator (27) with
one of  3 $H_{-4}\sim{H_{2}}$ scalar generators of $SO(2,d+2)$.
In particular, these generators have to satisfy
\eqn\grav{\eqalign{{\lbrack}L^{\beta\alpha},L^{+m}\rbrack
={\lbrack}L^{\beta{+}},L^{-m}\rbrack=0;\cr
{\lbrack}L^{\beta{-}},L^{+m}\rbrack=L^{\beta{+}};
{\lbrack}L^{\beta{+}},L^{-m}\rbrack=L^{\beta{m}};\cr
{\lbrack}L^{\beta\alpha},L^{++}\rbrack={\lbrack}L^{\beta{+}},L^{++}\rbrack
=0;\cr
{\lbrack}L^{\beta{-}},L^{++}\rbrack=L^{\beta{+}}
. }}
On the other hand, straightforward evaluation of the commutators of (28)
 with the generators of the SO(2,d) subalgebra gives
\eqn\grav{\eqalign{\lbrack{T},L^{-m}\rbrack
=\oint{{dz}\over{2i\pi}}
e^{-4\phi}\lbrace(\partial^2{X^m}\lambda-2\partial{X^m}\partial\lambda)
(\partial^2\varphi\lambda-2\partial\varphi\partial\lambda)\cr
+(\partial^2{X_n}\psi^n-2\partial{X_n}\partial\psi^n)(-\partial^2\varphi\psi^m
+2\partial\varphi\partial\psi^m)\rbrace;\cr
\lbrack{T,L^{++}}\rbrack=\lbrack{T,L^{+m}}\rbrack=0
.}}

Comparing (28) and (29) it is clear that the $T$-operator (27) must be
identified with $L^{\beta{+}}$. In addition, the comparison
of (28) and (29) also allows to deduce
 expressions for the $d$-vector generator of $H_{2}\sim{H_{-4}}$, namely,
\eqn\grav{\eqalign{
L^{\beta{m}}=\oint{{dz}\over{2i\pi}}
e^{-4\phi}\lbrace(\partial^2{X^m}\lambda-2\partial{X^m}\partial\lambda)
(\partial^2\varphi\lambda-2\partial\varphi\partial\lambda)\cr
+(\partial^2{X_n}\psi^n-2\partial{X_n}\partial\psi^n)(-\partial^2\varphi\psi^m
+2\partial\varphi\partial\psi^m)\rbrace}}

Next, evaluating the commutator
of $L^{\beta{m}}$ with the genarator $L^{-n}$ of the rotations in the
matter-Liouville planes, we deduce the second scalar generator $L^{\beta{-}}$:
\eqn\grav{\eqalign{
{\lbrack}L^{\beta{m}},L^{-n}\rbrack=\eta^{mn}L^{\beta{-}}\cr
L^{\beta{-}}=\oint{{dz}\over{2i\pi}}
e^{-4\phi}(\partial^2{X_l}\lambda-2\partial{X_l}\partial\lambda)
(2\partial\varphi\partial\psi^l-\partial^2\varphi\psi^l)
}}
It is now quite straightforward to obtain
the final, third scalar generator of $L^{\alpha\beta}$ of $H_2\sim{H_{-4}}$.
In particular it can be  deduced from the commutator
$\lbrack{L^{\beta{-}}},L^{\alpha{+}}\rbrack=-\eta^{+-}L^{\alpha\beta}$.
The computation is somewhat lengthy as  $L^{\beta{-}}$ is taken at picture 
$-4$ and $L^{\alpha{+}}$ at picture $-3$, so the obtained picture $-7$
generator must be transformed back to picture $-4$
by using the direct picture-changing operator
$:\Gamma:={i\over{{\sqrt{2}}}}(\psi_m\partial{X^m}+\lambda\partial\varphi)
+...$ (with insignificant $b-c$ ghost terms ignored).
The result is given by
\eqn\grav{\eqalign{
L^{\alpha\beta}={{i}\over{{\sqrt{2}}}}\oint{{dz}\over{2i\pi}}
e^{-4\phi}{\lbrace}{1\over4}(\partial\varphi)^5
-{3\over4}\partial\varphi(\partial^2\varphi)^2+{1\over4}
(\partial\varphi)^2\partial^3\varphi+
\lambda\partial\lambda(\partial^3\varphi-(\partial\varphi)^3)
\cr
-{3\over2}\lambda\partial^2\lambda\partial^2\varphi
+3\partial\lambda\partial^2\lambda\partial\varphi\rbrace
\equiv\oint{{dz}\over{2i\pi}}e^{-4\phi}F_2(\varphi,\lambda)\equiv
:(\oint{e^{-i\varphi}}\lambda)^3{\oint}e^{-4\phi+3i\varphi}\lambda:
}}
where, for  convenience, we have denoted the matter part of
$L^{\alpha\beta}$ by $F_2$
(c.f. $F_1$, the matter part of the $H_1\sim{H_{-3}}$
Cartan generator (19) of $SU(3)$).

Not surprisingly, 
this generator is just the $H_{-4}\sim{H_2}$ Cartan generator
of $SU(4)$ of the supersymmetric $c=1$ model ~{\selfdisc} with the
matter fields ${X}$ and $\psi$ of the $c=1$ theory replaced with
$\varphi$ and $\lambda$ 
(recall that we 
 neglect the effect of the dilaton field and the related
 background charge). 
As has been explained in ~{\selfdisc} this generator can be obtained
by taking the $H_{-4}$ generator $\oint{{dz}\over{2i\pi}}e^{-4\phi+3iX}\psi$
carrying the discrete $+3$ momentum and applying to it the SU(2)
lowering generator $\sim\oint{{dz}\over{2i\pi}}e^{-iX}\psi$ three times.
 This constitutes the complete set of the $H_{-4}\sim{H_2}$ generators.
It is now easy to compute the rest of the commutators
involving the $d+3$ generators of $H_{-4}\sim{H_2}$:
$(L^{\beta{m}},L^{\beta{+}},L^{\beta{-}}L^{\beta\alpha})$
of (27)-(32) and to show that, combined with
 $d+2$ generators $(L^{\alpha{m}},L^{\alpha{+}},L^{\alpha{-}})$
of $H_{-3}\sim{H_1}$ of (5),(8),(9) and ${{(d+1)(d+2)}\over2}$
isometry generators (12) (disregarding the Liouville dressing)
of $AdS_{d+1}$,  constitute
 ${{(d+3)(d+4)}\over2}$ generators  of $SO(2,d+2)$
satisfying the algebra (13), but with the capital indices
$M,N$ of (13) being now $d+4$-dimensional:
$M=(m,+,-,\alpha,\beta)$.
Therefore, just like in the case of $H_1\sim{H_{-3}}$,
the generators of $H_{2}\sim{H_{-4}}$ 
descend from the hidden extra dimension associeted with this
cohomology. 
The total number of extra dimensions (apart from the Liouville direction)
now equals $2$, in accordance with the concept relating 
 each ghost cohomology
(particularly, $H_1\sim{H_{-3}}$ and $H_2\sim{H_{-4}}$)
to an associate space-time dimension.
The remaining step is to show that the $H_2\sim{H_{-4}}$ generators
$L^{\beta{m}},L^{\beta{+}},L^{\beta{-}}$ and  $L^{\beta\alpha}$
of $SO(2,d+2)$ are indeed the space-time symmetry generators, i.e.
they induce $d+3$ higher order $\alpha$ symmetries
originating from the new space-time dimension associated with 
the higher order ghost cohomology $H_2\sim{H_{-4}}$.
 Using the $H_2\sim{H_{-4}}$ generators (27)-(32), it is
not difficult to  derive the space-time transformations
induced by (27) and (29) - (32).
We start from the first set of the $H_{2}\sim{H_{-4}}$ related
$\alpha$-transformations induced by $L^{\beta{+}}
=e^{2\phi}F_1(X,\psi)F_1(\varphi,\lambda)$.

Just like in the $H_{1}\sim{H_{-3}}$ case,
we choose to work with the positive picture $+2$ representation
of the $L^\beta$-operators ( the negative $-4$
picture represenation results in the equivalent set of transformations;
when applied to the RNS Lagrangian, the difference results
only in the total derivative terms).
Simple calculation shows that these transformations are given by

\eqn\grav{\eqalign{
\delta{X^n}=\epsilon^{\beta{+}}\lbrace
2e^{2\phi}\partial\psi^n
(\partial^2\varphi\lambda-2\partial\varphi\partial\lambda)
+\partial(e^{2\phi}\psi^n(\partial^2\varphi\lambda
-2\partial\varphi\partial\lambda))\rbrace\cr
\delta\psi^n=
\epsilon^{\beta{+}}\lbrace
-{e^{2\phi}}\partial^2{X^n}(\partial^2\varphi\lambda
-2\partial\varphi\partial\lambda)
-\partial({e^{2\phi}}\partial{X^n}(\partial^2\varphi\lambda
-2\partial\varphi\partial\lambda))\rbrace\cr
\delta\varphi=
\epsilon^{\beta{+}}\lbrace{e^{2\phi}}(\partial^2{X_m}\psi^m-2\partial{X_m}
\partial\psi^m)\partial\lambda
+\partial(e^{2\phi}\lambda(\partial^2{X_m}\psi^m-2\partial{X_m}
\partial\psi^m))\rbrace\cr
\delta\lambda=\epsilon^{\beta{+}}\lbrace
-{e^{2\phi}}(\partial^2{X_m}\psi^m-2\partial{X_m}
\partial\psi^m)\partial^2\varphi-2\partial(e^{2\phi}
(\partial^2{X_m}\psi^m-2\partial{X_m}
\partial\psi^m)\partial\varphi)\rbrace\cr
\delta\gamma={\epsilon^{\beta{+}}}e^{3\phi-\chi}{\lbrace}
2\partial\phi{F_1}(X,\psi)F_1(\varphi,\lambda)
+\partial(F_1(X,\psi)F_1(\varphi,\lambda))\rbrace\cr
\delta\beta=\delta{b}=\delta{c}=0}}
where $\epsilon^{\beta{+}}$ is  transformation parameter.
For technical reasons, it is also convenient to write explicitly 
the transformations for $\gamma$ in  terms of the bosonized fields
$\phi$ and $\chi$:
\eqn\grav{\eqalign{
\delta(\partial\phi)=2\epsilon^{\beta{+}}e^{2\phi}
F_1(X,\psi)F_1(\varphi,\lambda)\cr
\delta\chi=0}}
It is now easy to check that the transformations (33),(34)
leave the RNS action invariant, i.e. generate the global space-time symmetry.
Applying the transformations (33),(34) to the action (1)
and using some simple integration by parts
it is straightforward to show that

\eqn\grav{\eqalign{
\delta{S_{ghost}}={{\epsilon^{\beta{+}}}\over{\pi}}
\int{d^2z}(\bar\partial{e^{2\phi}})
F_1(X,\psi)F_1(\lambda,\varphi)\cr
\delta{S_{matter}}\equiv\delta(S_{X,\psi}+S_{Liouville})
=-{{\epsilon^{\beta{+}}}\over{{\pi}}}\int{d^2z}(\bar\partial{e^{2\phi}})
F_1(X,\psi)F_1(\varphi,\lambda)}}
so the $\alpha$-variation of the matter part is precisely cancelled by
that of the ghost part, just like in the $H_1\sim{H_{-3}}$ case.
The transformations (33),(34) thus constitute the first set of the higher order
$H_{2}\sim{H_{-4}}$ $\alpha$-symmetries. The second set, induced
by $L^{\beta{-}}$, is given by

\eqn\grav{\eqalign{
\delta{X^n}=\epsilon^{\beta{-}}\lbrace
2e^{2\phi}\partial\lambda(2\partial\varphi\partial\psi^n-\partial^2\varphi
\psi^n)
+\partial(e^{2\phi}\lambda
(2\partial\varphi\partial\psi^n-\partial^2\varphi\psi^n))\rbrace\cr
\delta\psi^n=\epsilon^{\beta{-}}\lbrace
e^{2\phi}(\partial^2{X^n}\lambda-2\partial{X^n}\partial\lambda)
\partial^2\varphi+2\partial(e^{2\phi}(\partial^2{X^n}\lambda-2\partial{X^n}
\partial\lambda)\partial\varphi)\rbrace\cr
\delta\varphi=-2\epsilon^{\beta{-}}\rbrace
e^{2\phi}(\partial^2{X_n}\lambda-2\partial{X_n}\partial\lambda)\partial\psi^n
-\partial(e^{2\phi}(\partial^2{X_n}\lambda
-2\partial{X_n}\partial\lambda)\psi^n)\rbrace\cr
\delta\lambda=\epsilon^{\beta{-}}\lbrace
-e^{2\phi}\partial^2{X_n}(2\partial\varphi\partial\psi^n-\partial^2\varphi
\psi^n)-2\partial(e^{2\phi}\partial{X_n}
(2\partial\varphi\partial\psi^n-\partial^2\varphi\psi^n))\rbrace\cr
\delta(\partial\phi)
=2{\epsilon^{\beta{-}}}
e^{2\phi}(\partial^2{X_n}\lambda-2\partial{X_n}\partial\lambda)
(2\partial\varphi\partial\psi^n-\partial^2\varphi
\psi^n)\cr
\delta\chi=\delta{b}=\delta{c}=0}}

The third set of $d$ space-time $\alpha$-transformations,
induced by the $d$-vector generator $L^{\beta{m}}$ of
$H_2\sim{H_{-4}}$, is given by

\eqn\grav{\eqalign{
\delta{X^n}=\epsilon^{\beta{n}}\lbrace
2{e^{2\phi}}\partial\lambda
(\partial^2\varphi\lambda-2\partial\varphi\partial\lambda)
+\partial(e^{2\phi}\lambda(\partial^2\varphi\lambda
-2\partial\varphi\partial\lambda))\rbrace\cr
+\epsilon^{\beta{m}}\lbrace
2e^{2\phi}\partial\psi^n(-\partial^2\varphi\psi_m
+2\partial\varphi\partial\psi_m)
+\partial(e^{2\phi}
\psi^n(-\partial^2\varphi\psi_m+2\partial\varphi\partial\psi_m))
\rbrace\cr
\delta\psi^n=\epsilon^{\beta{n}}\lbrace
e^{2\phi}\partial^2\varphi(\partial^2{X_m}\psi^m-2\partial{X_m}\partial\psi^m)
+2\partial(e^{2\phi}\partial\varphi(\partial^2{X_m}\psi^m-2\partial{X_m}
\partial\psi^m))\rbrace\cr
+\epsilon^{\beta{m}}\lbrace
e^{2\phi}
\partial^2{X^n}(-\partial^2\varphi\psi_m+2\partial\varphi\partial\psi_m)
+2\partial(e^{2\phi}\partial{X^n}
(-\partial^2\varphi\psi_m+2\partial\varphi\partial\psi_m))\rbrace\cr
\delta\varphi=\epsilon^{\beta{n}}\lbrace
{-}e^{2\phi}\partial\lambda
(\partial^2{X_n}\lambda-2\partial{X_n}\partial\lambda)
-\partial(e^{2\phi}\lambda(\partial^2{X_n}\lambda
-2\partial{X_n}\partial\lambda))\cr
+2{e^{2\phi}}\partial\psi_n
(\partial^2{X_m}\psi^m-2\partial{X_m}\partial\psi^m)
+\partial(e^{2\phi}\psi_n(\partial^2{X_m}\psi^m-2\partial{X_m}\partial\psi^m))
\rbrace\cr
\delta\lambda=\epsilon^{\beta{n}}\lbrace
-e^{2\phi}\partial^2\varphi
(\partial^2{X_n}\lambda-2\partial{X_n}\partial\lambda)
-2\partial(e^{2\phi}\partial\varphi
(\partial^2{X_n}\lambda-2\partial{X_n}\partial\lambda))\cr
+e^{2\phi}\partial^2{X_n}
(\partial^2\varphi\lambda-2\partial\varphi\partial\lambda)
+2\partial(e^{2\phi}\partial{X_n}
(\partial^2\varphi\lambda-2\partial\varphi\partial\lambda))\rbrace\cr
\delta(\partial\phi)=2\epsilon^{\beta{n}}
\lbrace(\partial^2{X_n}\lambda-2\partial{X_n}\partial\lambda)
(\partial^2\varphi\lambda-2\partial\varphi\partial\lambda)
\cr
+(\partial^2{X_m}\psi^m-2\partial{X_m}\partial\psi^m)
(-\partial^2\varphi\psi_n
+2\partial\varphi\partial\psi_n)\rbrace
\cr
\delta\chi=\delta{b}=\delta{c}=0}}

Finally, the transformations induced by $L^{\beta\alpha}$ are given by

\eqn\grav{\eqalign{
\delta\varphi=\epsilon^{\beta\alpha}\lbrace
-{5\over4}e^{2\phi}(\partial\varphi)^4
+{3\over4}e^{2\phi}(\partial^2\varphi)^2-{3\over2}
\partial(e^{2\phi}\partial\varphi\partial^2\varphi)-{3\over4}
e^{2\phi}(\partial\varphi)^2\partial^3\varphi-{1\over4}
\partial^2(e^{2\phi}(\partial\varphi)^2)\cr
+\partial^2(e^{2\phi}(\lambda\partial\lambda))-3e^{2\phi}
\lambda\partial\lambda(\partial\varphi)^2-{3\over2}
\partial(e^{2\phi}\lambda\partial^2\lambda)
-3e^{2\phi}\partial\lambda\partial^2\lambda\rbrace\cr
\delta\lambda=\epsilon^{\beta\alpha}\lbrace
\partial(e^{2\phi}\lambda(\partial^3\varphi-(\partial\varphi)^3))
+e^{2\phi}\partial\lambda(\partial^3\varphi-(\partial\varphi)^3)
+{3\over2}\partial^2(e^{2\phi}\lambda\partial^2\varphi)
\cr
-{3\over2}e^{2\phi}\partial^2\lambda\partial^2\varphi
-3\partial^2(e^{2\phi}\partial\lambda\partial\varphi)
-3\partial(e^{2\phi}\partial^2\lambda\partial\varphi)\rbrace\cr
\delta(\partial\phi)=2{\epsilon^{\beta\alpha}}e^{2\phi}F_2(\varphi,\lambda)\cr
\delta\chi=\delta{b}=\delta{c}=\delta{X^n}=\delta\psi^n=0}}

As before, $\epsilon^{\beta{-}},\epsilon^{\beta{n}},\epsilon^{\beta\alpha}$
are the transformation parameters related  to
$L^{\beta{-}},L^{\beta{n}}$ and $L^{\beta\alpha}$.
It is now straightforward to show that the 
RNS action (1) is invariant under the set of the
$\alpha$-transformations
(36) - (38). The proof is identical to the case of $L^{\beta{+}}$, 
demonstrated above. Similarly to 
the case of (33), (34) the variation of the matter 
part of the action (1) under the $\alpha$-transformations (36) - (38)
is cancelled by that of the superconformal ghost part.

To summarize, we have shown that the $RNS$ superstring action in $d$ dimensions
is invariant under the set of $d+3$ nonlinear space-time transformations 
induced by the generators of $H_2\sim{H_{-4}}$, in addition
to $d+2$ $\alpha$-symmetries of $H_1\sim{H_{-3}}$ and 
${1\over2}(d+1)(d+2)$   Poincare symmetries,
isomorphic to $SO({d},2)$ space-time isometries of $AdS_{d+1}$.
Thus the total space-time symmetry group is upgraded to
$SO(d+2,2)$, with one extra dimension contributed by 
generators of the ghost cohomology $H_1\sim{H_{-3}}$
and another by $H_2\sim{H_{-4}}$.
Note that, while the symmetries
induced by  $H_1\sim{H_{-3}}$ can be associated with those
of $2T$-physics, there is no obvious $2T$ counterpart
of the higher order symmetries of $H_2\sim{H_{-4}}$.
This concludes the proof that the generators of $H_{2}\sim{H_{-4}}$
effectively induce another new
 hidden space-time dimension, in addition to the one
produced by $H_1\sim{H_{-3}}$, in agreement with the concept 
of  `` a ghost cohomology = a hidden space-time dimension'', implying
that each 
 ghost
hohomology $H_{n}\sim{H_{-n-2}},n=1,2,3,...$
induces a set of space-time symmetries descending from the extra
dimension associated with $H_n$.

In the following section we will discuss the generalization of
our results to the general case of $n\geq{3}$.

\centerline{\bf 3. $H_{3}\sim{H_{-5}}$ and the Higher Cohomologies}

The construction of the $\alpha$-symmetry generators
 described above can be generalized to the 
case of higher ghost cohmologies as well, though manifest
expressions for the generators become more complicated.
As before, it is useful to take the operators
 $F_1(X,\psi)$ and $F_2(\varphi,\lambda)$ (26), (32) (structurally
related to the matter part of Cartan generators of $SU(3)$ 
and $SU(4)$ in the $c=1$ model)
as the building blocks to construct the currents of $H_{3}\sim{H_{-5}}$.
Since $F_1$ and $F_2$ are  primary fields of dimensions
${5\over2}$ and $5$ respectively, one has
$T(z)\partial{F_1(X,\psi)}(w)\sim{{{5{F_1(X,\psi)}}(w)}\over{(z-w)^3}}$ and
$T(z)\partial{F_2(\varphi,\lambda)}(w)\sim{{{10{F_2}}(w)}\over{(z-w)^3}}$
 and therefore $$R=
2\partial{F_1}(X,\psi)F_2(\varphi,\lambda)
-F_1(X,\psi)\partial{F_2(\varphi,\lambda)}$$ is 
a primary field of dimension ${{17}\over2}$.
Also $\Gamma(z):F_1(X,\psi):(w)\sim{O}((z-w)^{-2})$ and
$\Gamma(z):F_2(\varphi,\lambda):(w)\sim{O}((z-w)^{-3})$
since ${e^{-3\phi}}F_1$ and $e^{-4\phi}F_2$ are the elements
of $H_{-3}$ and $H_{-4}$ respectively.
Therefore, as the field contents of $F_1$ and $F_2$ do not mix
(one depends on $X$ and $\psi$ and another on the super Liouville mode),
$\Gamma(z)R(w)\sim{O}((z-w)^{-4})$.
Since $e^{-5\phi}$ has conformal dimension ${{15}\over2}$
and $\Gamma(z)e^{-5\phi}\sim{{O}((z-w)^5)}$ (disregarding the
irrelevant $c\partial\xi$ term of $\Gamma$),
the operator $\oint{{dz}\over{2i\pi}}e^{-5\phi}R$ 
is the integral of
primary field of dimension 1 annihilated by picture-changing,
i.e. the element of $H_{-5}$. This allows us to identify the first generator
of $H_{-5}\sim{H_3}$. Introducing the space-time 
index $\gamma$ associated with $H_{-5}\sim{H_{3}}$ (in addition to
$\alpha$ and $\beta$) we identify
\eqn\lowen{L^{\gamma{+}}\equiv{T}=
\oint{{dz}\over{2i\pi}}e^{-5\phi}(
2\partial{F_1}(X,\psi)F_2(\varphi,\lambda)
-F_1(X,\psi)\partial{F_2(\varphi,\lambda)})}
Using the same procedure as in the $H_{-4}\sim{H_2}$ case, 
it is not difficult to
construct the remaining $d+3$ currents generating the 
space-time $\alpha$-symmetries
on the level of $H_{-5}\sim{H_{3}}$:
\eqn\grav{\eqalign{
L^{\gamma{m}}={\lbrack}L^{\gamma{+}},L^{-m}\rbrack;m=0,...,{d-1}\cr
\eta^{mn}L^{\gamma{-}}={\lbrack}L^{\gamma{m}},L^{-n}\rbrack\cr
L^{\gamma\alpha}=::\Gamma^3:\lbrack{L^{\gamma{-}}},L^{\alpha{+}}\rbrack:\cr
L^{\gamma\beta}=::\Gamma^4\lbrack{L^{\gamma{-}}},L^{\beta{+}}\rbrack:
\equiv\oint{{dz}\over{2i\pi}}{e^{-5\phi}}F_3(\varphi,\lambda)}}
where the picture-changings in the last two commutators are necessary
to bring them back to the original picture $-5$.
As in the case of $H_{-4}\sim{H_{2}}$, the scalar
generator $L^{\gamma\beta}$ mixing the Liouville mode and the ghosts
(but commuting with $X$ and $\psi$) coincides
with the one of the Cartan generators of $c=1$ model -
this time, with the $H_{-5}\sim{H_3}$ generator of $SU(5)$,
with $F_3$ being its matter part.
As before, this generator can be obtained by the prescription
described in ~{\selfdisc}, namely, by applying the lowering
$SU(2)$ currents 4 times to $\oint{{dz}\over{2i\pi}}e^{-5\phi+4iX}\psi$:
\eqn\grav{\eqalign{
\oint{e^{-5\phi}}F_3(\varphi,\lambda)
\equiv:({\oint}e^{-i\varphi}\lambda)^4
\oint{e^{-5\phi+4i\varphi}\lambda}:}}
Alternatively, instead of starting with $L^{\gamma{+}}$ (which form
we simply have guessed) one could proceed in a more systematic way, 
deducing $L^{\gamma\beta}$ from the $H_{-5}\sim{H_{3}}$ Cartan
generator of SU(5) in $c=1$ model, replacing the $c=1$ matter fields
with $\varphi$ and $\lambda$ and then constructing other currents as 
the commutators of $L^{\gamma\beta}$ with the elements
of the previous cohomologies.
However, the latter procedure is  more complicated 
from the technical point of view, as
it involves more unpleasant picture changing transformations
and in addition the expressions for
the $SU(n)$ Cartan generators from cohomologies of high ghost numbers 
 are quite lengthy.
As before, $d+4$
 generators (39),(40) of $H_{-5}\sim{H_3}$ combined
with those of $H_{-4}\sim{H_2}$,$H_1\sim{H_{-3}}$ and
the Poincare generators of $SO(d,2)$, constitute
${{(d+3)(d+4)}\over{2}}$ generators of $SO(d+3,2)$.
These $d+4$ generators induce another set of higher order $\alpha$-symmetry
transformations at the level of $H_{-5}\sim{H_{3}}$, originating from another
hidden space-time dimension, labelled with the index $\gamma$.
By induction, this construction can be generalized to the cohomologies
$H_{n}\sim{H_{-n-2}}$ of arbitrary $n$ as well.
 In case of 
$n\geq{4}$, however, there seems to be no easy
way of deducing the form of the matter$+$Liouville generators
of $H_n\sim{H_{-n-2}}$ from the Cartan building blocks,
 as we did above with $L^{\alpha+}$ ,$L^{\beta{+}}$ and $L^{\gamma{+}}$.
So we have to start 
from the $H_n\sim{H_{-n-2}}$ Cartan generators of $SU(n+2)$.
Introducing the  index $\alpha_n$ for the extra dimension number $n$
associated with the generators of $H_n\sim{H_{-n-2}}$, we identify
the Cartan generator with 
\eqn\lowen{L^{\alpha_n\alpha_{n-1}}
=:(\int{{dz}\over{2i\pi}}e^{-i\varphi}\lambda)^{n+1}\oint{{dz}\over{2i\pi}}
e^{-(n+2)\phi+i(n+1)\varphi}\lambda:}
(again, we consider the special case of zero dilaton field so there is
no Liouville dressing).
As before, apart from $L^{\alpha_n\alpha_{n-1}}$ there are
$n$ additional scalar generators of $H_n\sim{H_{-n-2}}$, 
constructed as

\eqn\grav{\eqalign{L^{\alpha_n\alpha_i}=
\lbrack{L^{\alpha_n\alpha_{n-1}}},L^{\alpha_{n-1}\alpha_i}\rbrack,
i=1,...,n-2\cr
L^{\alpha_n\pm}=\lbrack{L^{\alpha_n\alpha_{n-1}}},L^{\alpha_{n-1}\pm}\rbrack
}}

and  one $d$-vector generator:

\eqn\lowen{L^{\alpha_n{m}}=\lbrack{L^{\alpha_n\alpha_{n-1}}}
,L^{\alpha_{n-1}m}\rbrack}
so  altogether there are $d+n+1$ generators in the $H_n\sim{H_{-n-2}}$
ghost cohomology, generating the $\alpha$-symmetry transformations
 associated with the $n$th hidden space-time dimension.
Combined with the generators of the lower cohomologies
$H_k\sim{H_{-k-2}}$ with $0\leq{k}\leq{n-1}$ these currents
would generate the $SO(d+n,2)$ space-time symmetry group.
Unfortunately, because of the complexity of the manifest expressions
for the currents of the higher cohomologies, we have not
been able to verify this explicitly for $n\geq{4}$.

\centerline{\bf 4. Conclusions}

The main result of this work shows that
 RNS superstring theories in various dimensions
possess a hierarchy of space-time symmetries
($\alpha$-symmetries), realized nonlinearly.
Eash class of $\alpha$-symmetries is generated by the currents
associated with  ghost cohomology $H_n\sim{H_{-n-2}}$ of number  $n$ .
By explicit construction, we have shown that each ghost cohomology
$H_{n}\sim{H_{-n-2}}$  has $d+n+1$ elements, namely,
$n+1$ scalar generators and one $SO(d-1,1)$ vector,
inducing $d+n+1$ space-time transformations. The RNS superstring action
is invariant under these transformations with the variation of the matter
part cancelled by that of the ghost part.
Combined together, the generators from  cohomologies
$H_k\sim{H_{-k-2}} (0\leq{k}\leq{n})$ generate the 
$SO(d+n,2)$ space-time symmetry group which is larger than
the ``naive'' SO(d,2) Poincare
 group of non-critical RNS superstring theory
in $d$ dimensions (i.e. the isometry group in the absence
of the $\alpha$-transformations mixing the matter and the ghost fields).
Thus each class of the $\alpha$-symmetries corresponding to 
$H_k\sim{H_{-k-2}}(1\leq{k}\leq{n})$ can be attributed to
 a hidden space-time dimension
(so altogether $n$ ghost cohomologies generate $n$ extra dimensions).
 In the simplest case of $n=1$,
the $\alpha$-symmetry generators on the level of $H_1\sim{H_{-3}}$
are in one to one correspondence to the generators
of the nonlinear space-time symmetries for a $AdS_d$ particle,
observed in the $2T$-formalism and which are also linked to a
hidden extra dimension in the Bars approach ~{\bars, \barss, \barsss}.
Our approach suggests, however, that the list of symmetries observed
in the $2T$ physics for a particle, is not complete as it corresponds
only to the lowest level $\alpha$-transformations of $H_1\sim{H_{-3}}$.
Therefore an interesting question is whether there is any interpretation
of the higher level $\alpha$-symmetries in the language of $2T$
physics. Operators from higher ghost cohomologies
, discussed in this paper, 
suggest that the number of hidden dimensions is bigger than $1$,
therefore the analogue of higher level $\alpha$-symmetries
should also exist for point particles, though these extra symmetries
have not yet been detected in the $2T$ approach.
In this work we have studied the hierarchy of $\alpha$-symmetries
on the classical level, i.e. as the space-time symmetry transformations of the
worldsheet RNS action. The important step forward would be to
generalize the discussion to the quantum level, in particular, to point
out the behaviour of the S-matrices and the
correlators under  the $\alpha$-transformations  and the related
conservation laws. In other words, what are the charges conserved as a result
of the $\alpha$-symmetries and what is their physical significance?
For example, in
the simplest case of the $c=1$ model the generators of $H_1\sim{H_{-3}}$,
enhancing the current algebra from the standard $SU(2)$ to $SU(3)$,
imply that  the tachyonic highest weight vectors
and their descendants at discrete momenta possess a new quantum number
interpreted as
``hypercharge'', conserved in interactions
of the ghost-dependent discrete states ~{\selfdisc}.
While the $c=1$ model is mainly an elegant
toy to play with (rather than a realistic phenomenological model),
so the ``hypercharge'' of discrete vertex operators
is hardly of any phenomenological significance,
the appearance of new non-trivial conservation laws related to the
interactions of ghost-dependent discrete states,
is by itself remarkable. One can ask if the conservation laws
associated with the $\alpha$-symmetries in higher dimensional
RNS superstring have any phenomenological interpretation.
It is possible that such an interpretation could lead to some
interesting stringy scenarios
of strong or even electroweak interactions.

In our paper we have investigated the limit of zero dilaton field,
in order to avoid complications related to the 
Liouville background charge.
In principle, it is straightforward to generalize
our discussion to include the effects of the dressing, though
 manifest expressions for the generators become more entangled,
except for the case of $d=9$ when the background charge is absent.
(alternatively, one can consider a critical string theory
in $d=10$ and compactify one of the dimensions on $S^1$).
The $d=9$ or the compactified $d=10$ cases are
 of the special interest since the generators
of $H_1\sim{H_{-3}}$ bring us immediately to $d=11$, relevant
to the $M$-theory dynamics while
`` switching on'' the higher cohomologies
such as $H_{2}\sim{H_{-4}}$ and $H_{3}\sim{H_{-5}}$  would 
further advance us to
the framework of $F$ and $S$-theories (compactified on a circle, if one
chooses to start from critical strings on $S^1$) ~{\barsf, \barsff}.
The problem is how the appearance of the higher
dimensions  can be explained dynamically. 
To address this question, one has to investigate the
worldsheet renormalization group flows, induced by the vertices
of nonzero cohomologies in the RNS sigma-model.
These RG flows are known to be stochastic, described by the Langevin-type
equations with the stochastic time, given  by the log of the 
worldsheet cutoff, that also plays the role of the
 extra dimension ~{\selfdisc}.
So far these RG flows have only been explored in the simplest
case of $H_1\sim{H_{-3}}$ operators with only one stochastic time
and one extra dimension present. As each cohomology contributes its own
associate hidden dimension, switching on the operators from higher
$H_n$'s naturally
 directs us to the concept of the stochastic processes
with multiple fictitious time variables.
It seems that no systematic understanding of such processes
exists at present, and this by itself
is of some interest.
Finally, another question for the future research
is related to the special case of critical 
RNS superstrings compactified on a circle of self-dual radius.
The case of the self-dual radius is peculiar because the number
of  space-time symmetry generators is larger: 
firstly, one still can build the 
$\alpha$-symmetry part of $SO(d+n,2)$ on the basis of 
$SU(n+2)$ Cartan generators of supersymmetric
$c=1$-model , as has been shown in this paper. 
These currents all carry momentum zero.
On the other hand, there are also the generators
with discrete momenta in the compactified direction,
 inheriting their  structure from all the $SU(n+2)$ generators
of the $c=1$ model, in addition to those 
rooted in the  Cartan subalgebra.
This enhanced space-time symmetry, appearing at the self-dual
compactification radius, needs a separate investigation
and may involve interesting relations between S
and T dualities, as well as phenomenological implications.

\centerline{\bf Appendix}

Here we present the details of the calculation showing
the absense of the Liouville-independent space-time vectors
in $H_2\sim{H_{-4}}$.
The most general expression for the vector generator of dimension 1
in the picture $-4$
is given by
\eqn\grav{\eqalign{V^n=
\alpha_1(\partial{X_m}\partial{X^m})(\psi_l\partial\psi^l)\partial{X^n}
+\alpha_2(\psi_m\partial{X^m})(\partial\psi_l\partial{X^m})\partial{X^n}
\cr
+\alpha_3(\psi_m\partial{X^m})(\psi_l\partial^2{X^l})\partial{X^n}
+\alpha_4(\partial{X_m}\partial{X^m})^2\partial{X^m}
\cr
+\beta_1(\partial{X_m}\partial{X^m})(\psi_l\partial\psi^l)\partial\psi^n
+\beta_2(\partial{X_m}\partial^2{X^m})\psi^n
+\beta_3(\partial{X_m}\partial{X^m})(\partial\psi_l\partial{X^l})\psi^n
\cr
+\beta_4(\partial{X_m}\partial{X^m})(\psi_l\partial^2{X^l})\psi^n
+\gamma_1(\psi_m\partial\psi^m)(\psi_l\partial{X^l})\partial\psi^n
+\gamma_2(\psi_m\partial^2\psi^m)(\psi_l\partial{X^l})\psi^n\cr
+\gamma_3(\psi_m\partial\psi^m)(\partial\psi_l\partial{X^l})\psi^n
+\gamma_4(\psi_m\partial\psi^m)(\psi_l\partial^2{X^l})\psi^n
+\lambda_1(\psi_m\partial^3\psi^m)\partial{X^n}\cr
+\lambda_2(\partial\psi_m\partial^2\psi^m)\partial{X^n}
+\lambda_3(\psi_m\partial^2\psi^m)\partial^2{X^n}+
\lambda_4(\psi_m\partial\psi^m)\partial^3{X^n}\cr
+\rho_1(\psi_m\partial{X^m})\partial^3\psi^n
+\rho_2(\psi_m\partial^2{X^m})\partial^2\psi^n
+\rho_3(\partial\psi_m\partial{X^m})\partial^2\psi^m
+\rho_4(\partial^2\psi^m\partial{X^m})\partial\psi^n\cr
+\rho_5(\partial\psi_m\partial^2{X^m})\partial\psi^n
+\rho_6(\psi_m\partial^3{X^m})\partial\psi^n
+\rho_7(\psi_m\partial^4{X^m})\psi^n
+\rho_8(\partial\psi_m\partial^3{X^m})\psi^n+\rho_9\cr
(\partial^2\psi_m\partial^2{X^m})\psi^n
+\rho_10(\partial^3\psi_m\partial{X^m})\psi^n
+\sigma_1\partial^5{X^n}+\sigma_2
(\partial{X_m}\partial^3{X^m})\partial{X^n}+\cr
\sigma_3
(\partial^2{X_m}\partial^2{X^m})\partial{X^n}+
\sigma_4(\partial{X_m}\partial^2{X^m})\partial^2{X^n}
+\sigma_5(\partial{X_m}\partial{X^m})\partial^3{X^n}}}
where $\alpha,\beta,\gamma,\lambda,\rho$ and $\sigma$
are some coefficients.
Computing OPE of $V^n$ with the stress tensor gives 22 primary
field constraints on $V^n$ following from the condition
that the OPE coefficients in front of all of the operators appearing
in terms
of the order of $(z-w)^{-n} (n\geq{3})$ must vanish,
i.e. the OPE has no singularities higher than quadratic.
That is, in our case, the most singular OPE term
(for generic $\alpha,\beta,\gamma,\lambda,\sigma,\rho$'s and $\delta_1$)
is of the order of $n=-6$.
Subsequently, the constraints for vanishing
of singularities of the order $n=-4,-5,-6$ give the 
first 9 linear equations of 22:

\eqn\grav{\eqalign{(12d+2)\sigma_2+8d\sigma_3+8\sigma_4+24\sigma_5
-18d\lambda_1-2d\lambda_2-18\rho_1-2\rho_3+2\rho_4
+18\rho_{10}-240\delta_1=0\cr
8\sigma_3+(4d+4)\sigma_4-4d\lambda_3-4\rho_2+4\rho_9-240\delta_1=0\cr
(4d+16)\alpha_4-d\alpha_1-\alpha_2-\beta_1+\beta_3-12\sigma_5-12\sigma_2=0\cr
4\alpha_3+(4d+4)\beta_2+8\beta_4+(8-4d)\gamma_2-48\rho_7-24\rho_{10}-24\rho_1
=0\cr
(2d+4)\alpha_1+2\alpha_2-\gamma_1+\gamma_3-12\lambda_4-30\lambda_1
+6\lambda_2=0\cr
2\alpha_2+(2d+4)\beta_1-d\gamma_1-12\rho_6-12\lambda_4-30\rho_1-6\rho_4=0\cr
(2d+4)\beta_3-d\gamma_3-12\rho_8-30\rho_{10}-6\rho_3=0\cr
2\alpha_2+2\beta_2+2d\beta_4+(2-d)\gamma_4
-6\rho_2-48\rho_7-6\rho_9=0\cr
2\sigma_2+2d\sigma_5-d\lambda_4-\rho_6+\rho_8-120\delta_1=0}}

The vanishing of the cubic terms gives the
remaining 13 primary field constraints:

\eqn\grav{\eqalign{4\sigma_4+12\sigma_5=0;
4\sigma_4+12\sigma_2+8\sigma_3=0;
4\rho_9+18\rho_{10}+2\rho_4=0\cr
2\rho_5+8\rho_9+12\rho_8=0;
2\rho_6+24\rho_7+2\rho_8=0;
8\rho_3+8\rho_4+4\rho_5=0\cr
8\rho_2+2\rho_5+12\rho_6=0;
18\rho_1+4\rho_2+2\rho_3=0;
18\lambda_1+2\lambda_2+4\lambda_3=0\cr
9\lambda_3+12\lambda_4=\beta_4+\beta_2=\delta_1=0\cr
2\beta_1+2\beta_3+2\gamma_1+6\gamma_2+2\gamma_3+4\gamma_4+2\lambda_2=0}}

The next set of constraints are the annihilation conditions by
$\Gamma$, $\Gamma{V}\sim{0}$. The analogous calculation
gives further $5$ constraints on the coefficients of (45),
following from the vanishing of all the OPE terms of the orders
of $(z-w)^{-n} (n=0,1,2)$ to ensure the annihilation by the direct 
picture changing:

\eqn\grav{\eqalign{
3\lambda_1+2\lambda_3+3\lambda_4+3\rho_1+2\rho_2+3\rho_6
+12(1-d)\rho_7-3d\rho_8-2d\rho_9-3d\rho_{10}=0\cr
\lambda_2-3\lambda_4+\rho_3-d\rho_4+(1-d)\rho_5-3d\rho_6+3\rho_8=0\cr
2\alpha_1+(d+2)\alpha_2+2(d-1)\alpha_3+2\beta_1+2\beta_2-6\lambda_1-
6\rho_1+6\sigma_2=0\cr
\alpha_1+\beta_1-2\beta_2-(d+2)\beta_3+2(d-1)\beta_4+6\sigma_5+6\rho_{10}=0\cr
\gamma_1+(1-d)\gamma_3+(4-2d)\gamma_4+6\lambda_4+6\rho_8=0}}

The constraints (46) - (48) are the necessary, but not the sufficient
cohomology conditions for $V$. In order to 
be the cohomology element generating global space-time 
symmetries, $V$ also has to be BRST non-trivial.
Conversely, to verify the BRST non-triviality, it is sufficient
to require that the action (1) is invariant under the space-time
transformations induced by the worldsheet integral of the
operator (45). The straightforward calculation
of the variations under
the space-time transformations induced by (45),
along with some partial integration,
shows that the invariance of the action leads to
the set of 29 further constraints on the coefficients of (45):

\eqn\grav{\eqalign{\rho_1-\rho_2+\rho_6-4\rho_7+\rho_8-\rho_9+\rho_{10}=0\cr
3\rho_1-2\rho_2-\rho_3+\rho_5+\rho_6-2\rho_8-2\rho_9=0\cr
3\rho_1-\rho_2-2\rho_3+\rho_4+\rho_5-4\rho_9=0\cr
\rho_1-\rho_3+\rho_4-\rho_7+\rho_8-\rho_9-2\rho_{10}=0\cr
-\rho_4+\rho_5-3\rho_6+\rho_8-2\rho_9+3\rho_{10}=0;
-\rho_4-\rho_5+\rho_6-3\rho_7+2\rho_8-\rho_9=0\cr
-\rho_2+\rho_3-2\rho_4+\rho_5-3\rho_9+3\rho_{10}=0;
-\rho_2-\rho_3-\rho_5+2\rho_6-3\rho_7+\rho_8=0\cr
-2\rho_1-\rho_2+\rho_3-\rho_4+\rho_6-\rho_7+\rho_{10}=0\cr
\rho_5=2\lambda_2-3\lambda_1=-3\lambda_2-\lambda_3+\lambda_4=-3\lambda_1
+3\lambda_2-2\lambda_3=0\cr
-3\lambda_1+\lambda_2+2\lambda_3-4\lambda_4=0;
\sigma_2-2\sigma_3-\sigma_4+2\sigma_5=0;
-\sigma_3-\sigma_4+2\sigma_5=0\cr
\sigma_2-\sigma_3-\sigma_4=0;
\sigma_2-\sigma_4=0\cr
2\beta_1-3\beta_2+2\beta_3-2\beta_4=0;
\beta_1-\beta_2+\beta_3-2\beta_4=0;\beta_1-\beta_2-\beta_4=0\cr
\beta_3-\beta_2-\beta_4=0;
\alpha_1=\alpha_2=\alpha_3=0;\gamma_1-3\gamma_2+\gamma_3-\gamma_4=0\cr
\gamma_1+2\gamma_2+\gamma_3-6\gamma_4=0;
\gamma_1+2\gamma_2-4\gamma_3-\gamma_4=0;
-4\gamma_1+2\gamma_2+2\gamma_3=0
}}
This is the set of 29 constraints for 31 coefficients
so the system has at least 2 independent nonzero solutions
(actually the number of independent solutions is larger since the
system is degenerate). Any choice of coefficients in (45)
satisfying (49) gives a 
space-time symmetry generator, so the number of the symmetry
generators is equal to the number of linearly independent solutions of
(49).
Now there are two possibilities: 
the first is that all these generators
differ only by BRST-trivial terms and are
 related to
the usual translation generator, transformed to picture $+2$.
This is the case if the solutions aren't compatible with
the  annihilation constraints  (48) (note that the picture $+2$
translation is by construction a primary field, so (46) and (47)
are satisfied automatically).
The second possibility is that the solutions of (49)
satisfy (46), (47) and (48). In this case the generator
is the element of ghost cohomology and induces
the $\alpha$-symmetry transformations on the level of $H_2\sim{H_{-4}}$.
Though we have described the case of 
generators at pictures $-4$ or $+2$, the same logic 
of search for $\alpha$-symmetries applies to case of
operators at higher ghost numbers as well, for this reason
we felt it would be instructive to demonstrate the above 
calculations in this appendix.

Since the overall number of constraints (46) - (49)
is bigger than the number of the coefficients in (45),
the appearance of the $\alpha$-symmetry is possible only in
case of the
degeneracy of the linear system of equations induced
by the primary, annihilation and symmetry constraints altogether.

It is not difficult to check 
that in the case under consideration ( picture $-4$ 
or $+2$ vector generators)
the constraints (46) - (49) have no nonzero solutions, therefore there are no
Liouville-independent generators of the $\alpha$-symmetry at this level.
The same can be shown for higher rank tensors as well.
This result ensures that we have no excessive
$\alpha$-generators and the set of the currents (27) and (30) - (32),
inducing the $\alpha$-symmetries at the level $H_2\sim{H_{-4}}$, is complete.

\listrefs
\end